\documentstyle[amssymb,epsfig,11pt]{article}
\newcommand{\be}{\begin{equation}}
\newcommand{\ee}{\end{equation}}

\def\ds{\displaystyle}
\def\ga{\gamma}
\def\th{\theta}

\def\p{\Phi}
\def\noi{\noindent}
\def\ol{\overline}
\def\d{\Delta}

\def\ga{\gamma}
\def\z{\zeta}
\def\b{\beta}
\def\pa{\partial}
\def\s{\sigma}
\def\t{\tilde}
\def\l{\lambda}
\def\ep{\epsilon}

\def\la{\langle}
\def\ra{\rangle}
\begin{document}
\title{Wandering of a contact-line at thermal equilibrium}

\author{Anusha Hazareesing, Marc M{\'e}zard}

\date{\today}
\maketitle

\begin{center}
Laboratoire de Physique Th{\'e}orique de l'Ecole Normale Sup{\'e}rieure
\footnote{Laboratoire propre du CNRS, associ\'e \`a l'ENS et \`a l'Universit\'e
de Paris XI}\\
24 rue Lhomond, 75231 Paris Cedex 05, France
\end{center}

\begin{abstract}
We reconsider the problem of the solid-liquid-vapour contact-line
on a disordered substrate, in the collective pinning regime. We go beyond
scaling arguments
and  perform an analytic computation, through the replica
variational method, of the fluctuations of the line. We show how gravity
effects must be
included for a proper quantitative comparison with available experimental data
of  the wetting of liquid helium on a caesium substrate.
The theoretical result is in good  agreement
with experimental findings for this case.
\end{abstract}

\section{Introduction}

When a liquid partially wets a solid, the liquid-vapour interface
terminates on the solid, at the contact line. If the solid surface
is smooth, then at equilibrium, we expect no distortions of the
contact line, and the Young's relation \cite{DG}
giving the contact angle in terms
of the interfacial tensions holds, that is

\be
\ga_{sv}-\ga_{sl}=\ga \cos(\th_{eq})
\ee

\noi where $\ga=\ga_{lv}$, and $\th_{eq}$ is the equilibrium mean contact
angle.

We consider a case where the substrate is weakly heterogeneous and where
the heterogeneities are ``wettable'' defects, leading
to a space dependance of the interfacial tensions
$\ga_{sv}$ and $\ga_{sl}$. Favoured configurations are
those where the liquid can spread on a maximum number of defects. We thus
expect distortions of the contact line which tends to be pinned by the
defects. Moreover, the energy due to the liquid-vapour interface induces
an elastic energy of the line. The competition between the elastic energy and
the pinning
due to the disorder gives rise to a non trivial wandering of the line, a
typical
example of the general problem of manifolds in random media \cite{FLN,HHZhang}.
The
case of the contact line is of special interest for several reasons.
There exists by now good experimental data for the correlations which
characterize the wandering of the line \cite{repain}. On the theoretical side,
the
problem presents two specific features. The elasticity of the line is
non local. The pinning energy due to
the surface heterogeneities is, up to a constant, a sum of local energy
contributions due to
the wetted defects. It has therefore non-local correlations which are of
the ``random field'' type in the usual nomenclature of manifolds in random
media.

In this paper we will consider the case of collective pinning where the
strength
of the individual pinning sites is small, but pinning occurs due to a
collective effect.
This  seems to be the relevant situation for the experiments. The case of
strong
pinning by individual impurities was studied by Joanny and De Gennes
\cite{DGJ}.
Collective pinning is a particularly interesting phenomenon since the balance
between
the elastic energy and the pinning one results in the existence of a special
length scale
$\xi$, first discussed by Larkin in the context of vortex lines in
superconductors
 \cite{Larkin}. This Larkin length is such that the lateral wandering of a
line,
thermalized  at low temperatures, on length scales smaller than
$\xi$, are less than the correlation length $\Delta$ of the disorder
(range of the impurities), while
beyond $\xi$ the lateral fluctuations become larger than $\Delta$ and
the line probes different impurities. The Larkin length scale diverges
in the limit where the strength of disorder goes to zero.
At zero temperature, the line has a single  equilibrium position when
its length is smaller than $\xi$, while metastable states appear only for
lengths larger than $\xi$. Therefore one can think of the contact line,
qualitatively, as an object which is rigid on small length scales (less than
$\xi$),
and fluctuates on larger length scales.
A third length scale which is relevant for the discussion is
the capillary length $L_c$, which is the length scale beyond which effects due
to gravity
become important: the line then becomes ``flat'' in the sense that its
fluctuations do
not grow any longer with the distance.

The collective pinning of the contact line was first addressed by Vannimenus
and Pomeau
\cite{VP}. They considered the case of very weak disorder in which the Larkin
length $\xi$
is larger than the capillary length. So their analysis only probes the ``Larkin
regime''
of length scales less than $\xi$, in which there exist only very few metastable
states.
A more complete qualitative picture, making clear the role of $\xi$, can be
obtained by
some scaling arguments originally developped for some related
problems by Larkin \cite{Larkin} and Imry-Ma \cite{IM}. For the case of the
contact line,
these arguments were introduced by Huse \cite{huse} and developped by De Gennes
\cite{DG}
and by Joanny and Robbins \cite{JR}. They lead to interesting predictions
concerning the
lateral fluctuations of the line: these should grow like the distance to the
power
$1/2$ on length scales less than $\xi$, and to the power $1/3$ on larger
distances
on length scales between $\xi$ and $L_c$.
More recently, Kardar and Ertaz \cite{KE} have performed a dynamic
renormalisation
group calculation for the contact-line at zero temperature,
subject to a uniform pulling force and
also find a roughness exponent $1/3$.
These scaling laws have been confirmed in recent experiments on the wetting
of helium on a caesium substrate \cite{repain}, confirming the validity of the
collective pinning picture in this case.

The aim of our paper is to go beyond the scaling analysis and provide a
quantitative computation
of the correlation function of the line.
We use the replica method together with a
gaussian variational approximation, with replica symmetry breaking \cite{MP}.
This approach, which is exact in the limit of large dimensions,  is known to
give good
results even for one dimensional systems as this one \cite{Engel,Kree,MPtoy}.
It confirms the scaling exponents  derived before, but also provides the
prefactors and a full description of the crossover between the two regimes
around the
Larkin length.

The paper is organised as follows:
We introduce the model in section (\ref{S1}).
In section (\ref{S2}), we present for completeness a scaling argument which
gives
the roughness
exponents, and we obtain an expression for the Larkin length by a
perturbative approach. In section (\ref{S3}), we present the  replica
calculation and
compute within a variational approximation the full correlation function
in the limit of low temperatures, first neglecting gravity effects, and then
including them.
In section  (\ref{S4}), we compare  our theoretical prediction with
experimental data.

\section{The model}\label{S1}

\begin{figure}[h,t]
$$
\epsfig{file=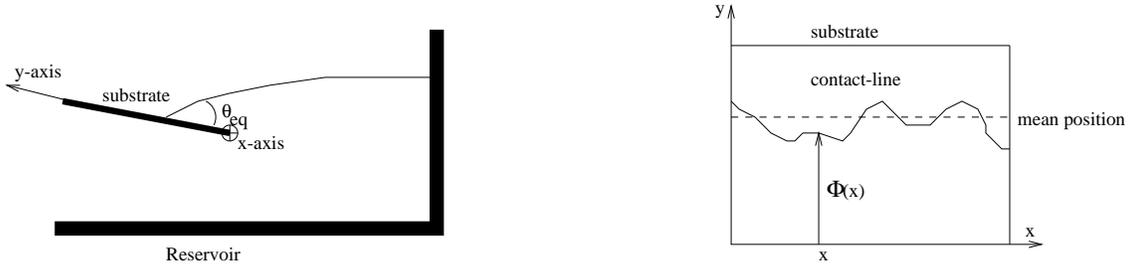,width=15cm}
$$
\caption{Sketch of the experimental set up}
\label{figure1}
\end{figure}

Consider a situation given by figure (\ref{figure1}), where the liquid wets
an impure substrate which
is slightly inclined with respect to the horizontal. We denote by $(x,y)$ the
space
co-ordinates of the substrate. The excess energy per unit area  due to pinning
is given by

\be\label{en_defaut}
e(x,y)=\ga_{sl}(x,y)-\ga_{sv}(x,y)-\ol {\ga_{sl}(x,y)-\ga_{sv}(x,y)}
\ee

\noi resulting in a total pinning energy

\be\label{en_pinning}
\int_{0}^{L} dx \int_{0}^{\p(x)} dy \  e(x,y)
\ee

\noi where $\p(x)$ is the height  of the the contact line at the
abscissa $x$ (overhangs are neglected),  and $L$ the width of the substrate.
As for the pinning energy per unit area or force per unit length
$e(x,y)$, we shall suppose that it is gaussian
distributed,  which is the case if it results from a large number of
microscopic
interactions, and that it has local correlations on length scales of order
$\d$. Specifically, we choose

\be\label{corr_en}
\ol {e(x,y)e(x',y')} = \frac{W}{\d^2} \delta (x-x')  C \biggl(\ds \biggl|
\frac{y-y'}{\d} \biggr|  \biggr)
\ee
\noi where the correlation function $C(r)$ is normalised to $C(0)=1$
and $C''(0)=-1$, and decreases
fast enough to zero for $r \gg 1$. The asymmetry introduced in (\ref{corr_en})
between the two directions $x$ and $y$ is for computational convenience.
In most physical situations,the distribution of disorder should be isotropic
in the $x-y$ plane, leading to a correlation in the $x$ direction on length
scales of order $\d$. As we shall explain below, we have found
 that  this correlation has only small effects,
which exist only on very short distances and are not relevant experimentally.
As for the shape of the function $C(r)$, we shall first use for simplicity

\be\label{function_f}
C(r)=f(r)=\exp (-r^2/2) \ ,
\ee

\noi  and we shall later comment on the modification of our result for more
general correlations.

We must also add to the random potential term, a capillary energy term,
which, if we neglect gravity and suppose
that the slope of the liquid-vapour interface varies smoothly, is given by
\be\label{en_cap}
E_{\rm cap}=\frac{c}{2} \ds \int_{\frac{2\pi}{L} \leq |k| \leq \frac{2\pi}{\d}}
\frac{dk}{2\pi} |k||\p(k)|^2
\ee
\noi  with $c=\ds \frac{\ga \sin ^{2}\th_{eq}}{2}$, $\th_{eq}$ being the
average
equilibrium contact angle\cite{JR}.

\noi The final hamiltonian is thus
\be\label{eq1}
H=\frac{c}{2} \ds \int_{\frac{2\pi}{L} \leq |k| \leq \frac{2\pi}{\d}}
\frac{dk}{2\pi} |k| \ |\p(k)|^2 +
\int_{0}^{L} dx\ V(x,\p(x))
\ee
\noi where $\ds V(x,\p)=\int_{0}^{\p(x)} dy \  e(x,y)$. As a sum of independent
gaussian
variables, $V(x,\p)$ is gaussian distributed, and up to a uniform
arbitrary random shift
we can choose:

\be\label{eq2}
\ol {V(x,\p)V(x',\p')}=-W\delta (x-x')f
\biggl(\biggl(\frac{\p-\p'}{\d}\biggr)^2\biggr)
\ee

\noi where $f(u)$ is a function which grows as $\sqrt{|u|}$ for large $|u|$.
Its precise form depends on the correlation function $C$ of the energy per unit
area,
and is given in the simple case (\ref{function_f}) by

\be\label{eq2bis}
f(u^2)=\ds |u|\int_{0}^{|u|}dv\ e^{-v^2/2} - (1-e^{-u^2/2})
\ee

This model provides a good description of the problem of a contact
line on a disordered substrate under the following hypotheses:

$\bullet$
\noi The slope of the liquid-vapour interface is everywhere small.

$\bullet$
\noi The length of the contact line is small compared with the capillary
length, so that one can neglect gravity.

$\bullet$
\noi The defects in the substrate are weak and give rise to collective
pinning.

The main part of our work will be dedicated to the analytical study of this
simplified model.
We shall then examine the corrections due to gravity, to more general
correlations of
the disorder, and to the correlations in the $x$ direction.
The quantity which is measured experimentally and which we shall compute is the
correlation
function of
the  position of the line:
\be
\ol{ \la (\Phi(x)-\Phi(y))^2 \ra}
\ee
where  thermal averages are denoted by angular brackets and the
average over disorder by an overbar. As we shall see, in different length
regimes, this
correlation increases as a power law, which defines locally the wandering
exponent $\zeta$ from:

\be
\ol{ \la (\Phi(x)-\Phi(x))^2 \ra} \sim |x-x'|^{2 \zeta} \ .
\ee

\section{Perturbation theory and scaling  arguments}\label{S2}
For completeness we rederive in this section an expression for the Larkin
length
by perturbation theory, and review the scaling derivation of  the roughness
exponents.

\subsection{The Larkin length}

\noi On a sufficiently small length scale, we can assume  that the difference
in heights between any two points is small compared with the correlation length
$\d$
of the potential. We can thus linearize the potential term \cite{Larkin}
such  that $V(x,\p (x)) \simeq V(x,0)-e(x)\p (x)$. This leads to a random force
problem with a force correlation function
$\ds \ol {e(x)e(x')} = \frac{W}{\d^2} \delta (x-x')$.
Rewriting the hamiltonian as:

\be\label{eq3}
\ds H =\ds  \frac{c}{2} \ds \int \frac{dk}{2\pi}\ |k| \
\biggl|\p(k)-\frac{e(k)}{c|k|}
\biggr|^2
- \frac{1}{2c} \ds \int \frac{dk}{2\pi}\ \frac{|e(k)|^2}{|k|}
\ee

\noi we get for $T \rightarrow 0$ and $\d \ll |x-x'| \ll L$,

\be
\ol {<(\p(x)-\p(x'))^2>}=\frac{2W}{c^2\d^2}\int \ \frac{dk}{2\pi}
\frac{(1-\cos(k(x-x')))}
{k^2}
=\frac{W}{c^2\d^2}|x-x'|
\ee

\noi The wandering exponent in the Larkin regime is given by $\zeta= 1/2$.
The linear approximation is no longer valid
when $|\p(x)-\p(x')|$ becomes
of the order $\d$. Typically $|x-x'|$ is then of order
$\ds \xi = \frac{c^2\d^4}{W}$,
where $\xi$ is the so called  Larkin length.

\subsection{Roughness exponent for large fluctuations}

On length scales larger than $\xi$, the fluctuations of the
line are greater than the correlation length $\d$ and perturbation theory
breaks down. One can estimate the wandering exponent by a simple scaling
argument
as follows \cite{MP}.
The hamiltonian is given by (\ref{eq1}) and  we can no longer linearize the
potential term in (\ref{eq1}).

We consider the scale transformation, $x \rightarrow lx$,
$\p(x) \rightarrow  l^{\z}\p(x)$, $V(x,\p(x)) \rightarrow
l^{\lambda}V(x,\p(x))$.
Imposing that the two terms in the hamitonian scale in the same way
and that the potential term keeps the same statistics after rescaling, we
have

\be
\lambda = 2\z -1 \quad \mbox{and} \quad 2\lambda = -1 + \z
\ee

\noi and so $\z = 1/3$. Note that this is less than the value $1/2$
obtained in the Larkin regime.
On a still larger length scale (larger than the capillary length),
we expect the line to be flat and $\z = 0$.

This exponent can be recovered by the following Imry-Ma argument
\cite{IM,huse,DG,JR}.
On a scale $L$, the line fluctuates over a distance  $\p$. The elastic energy
contribution
then scales as $c{\p}^2$.
As for the pinning energy, since it is a sum of independant gaussian variables,
it scales  as $\ds \sqrt{W\d}\sqrt{\frac{L\p}{\d ^2}}$,
where $\sqrt{W\d}$ is a measure of
the pinning energy on an area $\d ^2$ and
$\ds \frac{L\p}{\d ^2}$ an order of magnitude of the number of such pinning
sites.
Minimising the total energy $\ds c{\p}^2-\sqrt{W\d}\sqrt{\frac{L\p}{\d ^2}}$
with respect
to $\p$, we get $\ds \p \simeq  \d \biggl(\frac{L}{\xi}\biggr)^{1/3}$
with $\ds \xi  \sim  \frac{c^2\d^4}{W}$.

\section{The replica computation}\label{S3}
\subsection{Computation of the free energy}

We now turn to a microscopic computation of  the free energy $F=-T\ln Z$.
Since the free energy is a self averaging quantity, the typical free energy is
equal to the average of $F$ over the disorder. We compute it from the replica
method with  an
analytic continuation of $\ol {Z^n}$, for $n \rightarrow 0$ \cite{MPV}.
The $n^{th}$ power of  the partition function

\begin{eqnarray}
Z^{n}=\int \prod_{a=1}^{n} d[\p_{a}]\ \exp \left\{-\frac{\b c}{2}\int
\frac{dk}{2\pi} \sum_a |k||\p_a(k)|^2 -\b \sum_a \int_{0}^{L}dx
\ V(x,\p_a(x))\right\}
\end{eqnarray}

\noi gives after averaging over the disorder

\begin{eqnarray}
\ol {Z^n}=\int \prod_{a=1}^{n} d[\p_{a}]\ \exp \left\{-\b {\cal H}_{n}
[\p_{a}]\right\}
\end{eqnarray}

\noi where

\be\label{eq6}
{\cal H}_{n}=\frac{c}{2}\int \frac{dk}{2\pi} \sum_a |k||\p_a(k)|^2
+ \frac{\b W}{2} \sum_{a,b} \int_{0}^{L}dx \
f \biggl(\biggl(\frac{\p_{a}(x)-\p_{b}(x)}{\d}\biggr)^2\biggr)
\ee

\noi We note that the expression of the free energy is
invariant with respect to a translation of the centre of mass of the
line $\p_{CM}=\frac{1}{L}\int_{0}^{L}dx\ \p(x)=\frac{1}{L}\p(k=0)$.
We can fix the centre of mass so that  there is no integration on the $k=0$
mode.
The partition function $\ol {Z^n}$ cannot be computed  directly.
Following \cite{MP}, we perform a variational calculation  based on the
variational hamiltonian

\be
{\cal H}_{o}=\int \frac{dk}{2\pi} \sum_{a,b=1}^{n} \p_a(-k)G_{ab}^{-1}(k)
\p_{b}(k)
\ee

\noi where $G^{-1}$ is a hierarchical Parisi matrix.

\noi The variational free energy

\be
{\cal F}=\frac{-1}{\b n}\ln Z_{o}+\frac{1}{n}<{\cal H}_{n}-{\cal H}_{o}>_{o}
\ee

\noi gives up to a constant term,

\be\label{eq8}
\frac{{\cal F}}{L}=\lim_{n \rightarrow 0}\ \frac{1}{n}
\biggl[\frac{-1}{2\b}\int \frac{dk}{2\pi} \mbox{Tr}_{a}\ln G+
\frac{c}{2\b}\int \frac{dk}{2\pi}
|k|\sum_{a} G_{aa}(k)+\frac{\b W}{2}\sum_{a \not = b}{\hat f}
\biggl(\frac{B_{ab}}{\d^2}\biggr)\biggr]
\ee

\noi where

\be
\hat{f} (z)  \  =\int_{-\infty}^{\infty}\ \frac{du}{\sqrt{2\pi}}
f(u^2 z)\ e^{-u^2/2}\ = \sqrt{1+z}-1
\ee

\noi and

\be
B_{ab}=\frac{1}{\b} \int \frac{dk}{2\pi} \
(G_{aa}(k)+G_{bb}(k)-2G_{ab}(k))
\ee

\noi The optimal free energy is
obtained for a matrix $G$ verifying the stationarity conditions
$\ds \frac{\pa {\cal F}}{\pa G_{ab}}=0$, which read

\be\label{eq9}
\begin{array}{l}
G_{ab}^{-1}=\ds \frac{-2\b W}{\d^2}{\hat f'}\biggl(\frac{B_{ab}}{\d^2}\biggr)
\quad
\mbox{for} \quad a \not = b \\ \\
\ds \sum_{b}G_{ab}^{-1}=c|k|
\end{array}
\ee

\noi More details on this approach can be found in \cite{MP,MPV,BMP}.

\subsection{The replica symmetry breaking solution}

To solve equations (\ref{eq9}), we suppose that the matrix $G$ has a
hierarchical
replica symmetry breaking stucture {\it \`a la Parisi}. We can write
$G_{ab}^{-1}=(c|k|-{\t \s})\delta_{ab}-\s _{ab}$. $G^{-1}$ is thus parametrised
by
a diagonal part $c|k|-{\t \s}$, and a function $\s (u)$
defined on the interval $[0,1]$
\footnote{From equations (\ref{eq9}), we
know that the off-diagonal elements of $G^{-1}$ do not
depend on $k$.}.
The optimisation equations for $G$ can then be written as:

\be\label{eq10}
\s (u)=\frac{2\b W}{\d^2}{\hat f'}\biggl(\frac{B(u)}{\d^2}\biggr)
\ee

\noi with

\be\label{eq11}
B(u)=\frac{2}{\b}\int \frac{dk}{2\pi} ({\t g(k)}-g(k,u))
\ee

\noi  The solution to these equations is described in appendix {\bf A}.
It is best written
in terms of the function

\be
[\s ](u)=u\s (u)-\int_{0}^{u} \ dv\ \s(v)
\ee

\noi which is given by

\be\label{eq12}
\begin{array}{l}
[\s ](u)=\ds \frac{W}{\pi c \d^4}\ds \biggl(\frac{u}{u_c}\biggr)^{3/2}
\quad   \mbox{for} \quad u \leq u_{c}

\\ \\

[\s ](u)=\ds \frac{W}{\pi c \d^4} \quad  \mbox{for} \quad u \geq u_{c}
\end{array}
\ee

\noi We shall give the value of the breakpoint $u_c$ in the (experimentally
relevant)
limit of low temperatures. Defining

\be\label{temp}
T_c= \pi c \frac{\Delta^2}{3}\ ,
\ee

\noi we get  $u_c\simeq T/T_c$. From the expression

\be
G_{aa}(k)=\frac{1}{c|k|}\biggl[1+\int_{0}^{1}\frac{du}{u^2}
\frac{[\s ](u)}{[\s ](u)+c|k|}\biggr]
\ee

\noi we obtain in the regime
$\Delta \ll |x-x'| \ll L $

\be\label{corr}
\sqrt{\ol {<(\p(x)-\p(x'))^2>}} =\d{\cal H}\biggl(\frac{x-x'}{\xi} \biggr)
\ee

\noi  where

\be\label{asymp}
{\cal H}^2(x) =\frac{4}{3} \biggl(\frac{x}{\pi}\biggr)^{2/3}
\int_{0}^{\infty}dk
\frac{(1-\cos(k))}{k^{5/3}} \int_{0}^{(x/{\pi k})^{1/3}}\frac{dw}{w^3+1}\
+ \frac{2x}{3\pi} \int_{0}^{\infty}dk \frac{(1-\cos(k))}{k(x/{\pi}+k)}
\ee

\noi and $\ds \xi=\ds \frac{c^2 \d^4}{W}$.

\begin{figure}[h,t]
$$
\epsfig{file=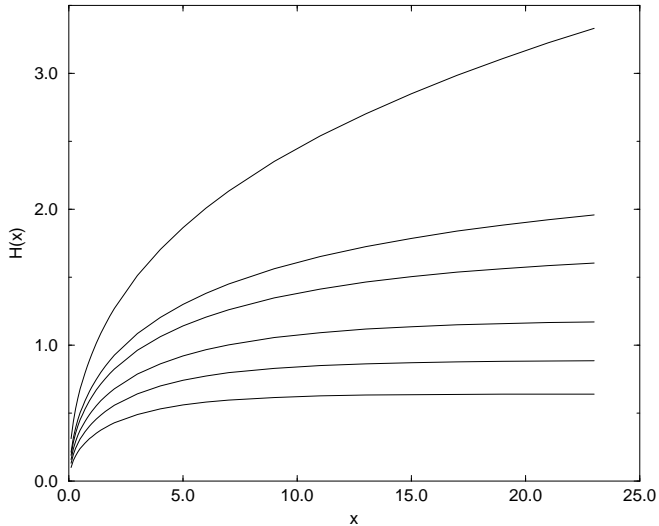,width=8.5cm,angle=-90}
$$
\caption{The theoretical predictions for the rescaled height correlations.
{}From top to bottom, the curve ${\cal H}(x)$  (\ref{corr}) where gravity is
neglected
($\l= \pi \xi/L_c=0$), and the curves
${\cal H}_g(x,\l)$ for values of $\l$ such that
$\l=0.1,0.2,0.4,0.6,0.8$. Notice the need to have a very small ratio of Larkin
length to capillary length in order for the zero gravity result to be good.}
\label{fit1}
\end{figure}

\noi The function ${\cal H}$ is the analytical prediction in the regime where
gravity
effects can be neglected.
It is plotted in figure \ref{fit1} and has the following asymptotic behaviour.
For small $x$, ${\cal H}(x) \simeq \sqrt{|x|}$ and for large $x$,
${\cal H}(x) \simeq 1.14 |x|^{1/3}$. Therefore the predictions in the various
scaling
regimes, including prefactors, is as follows.

\noi When $|x-x'| \ll \xi$, (Larkin regime):

\be\label{asymp1}
\sqrt{\ol {<(\p(x)-\p(x'))^2>}} \simeq  \d  \left|\frac{x-x'}{\xi}\right|^{1/2}
\ee

\noi  When $|x-x'| \gg \xi$, (random manifold regime):

\be\label{asymp2}
\sqrt{\ol {<(\p(x)-\p(x'))^2>}} \simeq 1.14 \d
\left|\frac{x-x'}{\xi}\right|^{1/3}
\ee

\noi Before turning to the comparison with the experiment, we first compute
various
correction factors to this formula.

\subsection{A more general form of  the disorder}
We show that even in the more general case where we only impose
that the correlation function of the potential has the asymptotic behaviour
$f(|u|) \sim \sqrt{|u|}$ for large $|u|$, the height correlation function
can be put in the form

\be\label{corr_general}
\sqrt{\ol {<(\p(x)-\p(x'))^2>}} =\d {\cal G}\biggl(\frac{x-x'}{\xi}\biggr)
\label{corrgen}
\ee

\noi in the limit $T \rightarrow 0$, and for $\d \ll |x-x'| \ll L$.
The derivation of ${\cal G}$, for an arbitrary correlation
function $f$, is given in appendix {\bf B}.

\subsection{Effect of the cut-off $\d$ at small $x$}\label{sub_delta}

Coarse-graining the force in the $x$ direction on scales of order $\d$ leads
to a discretized (in $x$) version of (\ref{corr_en}), which is equivalent to
the form which we use, but with a cut off of order $\d$ at small $x$.
On scales comparable to $\d$, there are thus
corrections to equation (\ref{asymp}) due to this cutoff, which are easily
computed.
When we take into account the cut-off $\d$, the correlation function
${\cal H}$ becomes

\be\label{asymp_delta}
{\cal H}_{1}^2(x,\nu) =\frac{4}{3} \biggl(\frac{x}{\pi}\biggr)^{2/3}
\int_{0}^{2\pi \nu x}\frac{dk}{k^{5/3}}
(1-\cos(k)) \int_{0}^{(x/{\pi k})^{1/3}}\frac{dw}{w^3+1}
\ + \frac{2x}{3\pi} \int_{0}^{2\pi \nu x}\frac{dk}{k(x/{\pi}+k)} (1-\cos(k))
\ee

\noi where $\ds \nu = \frac{\xi}{\d}$. The  effect of the cut-off is to shift
slightly
downwards the theoretical curve, especially in the region of small $x$.

\subsection{Effect of gravity}

In the geometry considered, the effective
capillary length is given by $\ds \sqrt{\frac{\ga}{\rho g \sin \alpha}}$,
where $\alpha$ is the tilt angle of the substrate with respect to
to horizontal \cite{repain}.
To take into account gravity we must replace the kernel $|k|$ in the
hamiltonian
(\ref{eq1}) by $\sqrt{k^2+1/L_c^2}$,  where $L_c$
is the capillary length. Generalising the previous calculation we
have computed the corrections due to gravity. The fluctuation of the line at
distance $x$ now
depends on two parameters, $x/\xi$ and $\l=\pi \xi/L_c$. The computation done
in appendix
{\bf C} expresses  $[\s](u)$
in the terms of an inverse function ${\cal I}^{-1}$. When the Larkin length is
sufficiently
small compared with the capillary length, that is  when
$\xi$ is less than approximately  $0.5 L_c$,
we have in the limit where $T \ll T_c$

\be
\begin{array}{l}
[\s ](u)=0 \quad   \mbox{for} \quad u \leq u_1

\\ \\

[\s ](u)=\ds \mu c \ {\cal I}^{-1}
\biggl(\frac{L_c}{\pi  \xi}\ds  \biggl(\frac{u}{u_c}\biggr)^{3/2} \biggr)
\quad   \mbox{for} \quad u_1 \leq u \leq u'_{c}

\\ \\

[\s ](u)=\ds \mu c \ {\cal I}^{-1}
\biggl(\frac{L_c}{\pi  \xi}\biggl(\frac{u'_c}{u_c}\biggr)^{3/2} \biggr)
\quad  \mbox{for} \quad u \geq u'_{c}

\\ \\

\mbox{with} \quad \ds u_1 =u_c \biggl(\frac{\pi  \xi}{L_c} {\cal
I}(0)\biggr)^{2/3}
\quad \mbox{and}  \quad
u_{c} \simeq \ds \frac{T}{T_c}
\end{array}
\ee

\noi As for $u'_c$, it is slightly larger than $u_c$ and also of order $\ds
\frac{T}{T_c}$.
 When the capillary length
goes to infinity $u_1$ tends towards $0$ and $u'_c$ towards $u_c$.
\noi The function ${\cal I}(x)$ is given by

\be
{\cal I}(x)=\ds \biggl(\int_{0}^{\infty}\frac{dk}{(\sqrt{q^2+1}+x)^2}\biggr)^2
\biggl(2\int_{0}^{\infty}\frac{dk}{(\sqrt{q^2+1}+x)^3}\biggr)^{-3/2}
\ee

\noi and for large $x$, ${\cal I}(x) \simeq x$. This asymptotic behaviour
ensures
that we do recover the results of section (\ref{S4}) when $\mu$ goes to zero.
The correlation is now  given by

\be\label{corr_grav}
\sqrt{\ol {<(\p(x)-\p(x'))^2>}} =\d{\cal H}_g\biggl(\frac{x-x'}{\xi},
\frac{\pi  \xi}{L_c} \biggr)
\ee

\noi where

\be
\begin{array}{ll}
{\cal H}^2_g(x,\l)=&\ds \frac{4}{3} \int_{0}^{\infty} dk
\frac{(1-\cos(kx/\pi)}{\sqrt{k^2+\l^2}}
\ds \int_{(\l {\cal I}(0))^{1/3}}^{(u'_c/u_c)^{1/2}}\frac{dw}{w^3}
\frac{\l {\cal I}^{-1}(w^3/\l)}{\l {\cal I}^{-1}(w^3/\l)+\sqrt{k^2+\l^2}}

\\ \\

& + \ds \frac{2}{3} \frac{u_c}{u'_c}  \int_{0}^{\infty} dk \
\frac{(1-\cos(kx/\pi)}{\sqrt{k^2+\l^2}}
\frac{\l {\cal I}^{-1}(1/\l)}{\l {\cal I}^{-1}(1/\l)+\sqrt{k^2+\l^2}}
\end{array}
\ee

\noi The asymptotic behaviour of ${\cal H}_g$ is different from that of $\cal
H$.
For small $x$

\be
\ds {\cal H}^2_g(x,\l) \simeq x \biggl \{
\frac{2}{3} \int_{(\l {\cal I}(0))^{1/3}}^{(u'_c/u_c)^{1/2}} dw \
\frac{\l}{w^3}{\cal I}^{-1}(w^3/\l) + \frac{u_c}{3u'_c} \l {\cal I}^{-1}(1/\l)
\biggl \}
\ee

\noi and for large $x, {\cal H}_g(x,\l)$ tends towards a constant depending on
$\l$.

\noi From the previous equations we can see that gravity has a significant
effect when
$\ds \l=\frac{\pi  \xi}{L_c}$ becomes of order $1$, where $\xi$ is the Larkin
length
and $L_c$ the capillary length. Moreover we can also note that the correction
for small $\l$ to the case without gravity is of order $\l ^{1/3}$. The limit
$\l$ going to $0$ is thus a rather slow one. This is illustrated  in figure
(\ref{fit1}).

\section{Comparison with experiment}\label{S4}

\subsection{The experimental set up}
We have fitted the data from experiments carried out by C.Guthmann and
E.Rolley \cite{repain} with our theoretical curve.
The experiments study the wetting properties of
liquid helium 4 on caesium below the wetting transition temperature which is
about $2K$. Above that temperature caesium is wetted by helium.
In the experiments carried out by Guthmann and Rolley  the substrate consists
of caesium
deposited on a gold mirror which is slightly inclined with respect to the
horizontal
(see figure \ref{figure1}).  The wetted defects are small areas on the
substrate where the caesium has been oxydised.
The experiments are carried out on a range of temperatures going from about
$1K$ to $2K$.
There is a constant inflow of helium at the bottom of the helium reservoir
to maintain the contact angle to its maximum value $\th_a$,
the advancing angle, which is in general different from the
equilibrium contact angle $\th_{eq}$ (see figure \ref{figure1}).
This is necessary because otherwise, the liquid would recede and
the contact angle would  shrink to zero due to strong hysteresis.
Height correlations are calculated from snapshots  of the advancing line
when it is pinned. The incoming  helium is regulated to ensure that the line
moves with  a small velocity and so we can  probably suppose can we are just at
the
limit of depinning.

The predicted order of magnitude of $T_c$ given by (\ref{temp}) is
$\ds \frac{\pi c \d^2}{3} = \ds \frac{\pi \ga \sin^2 (\th_{eq}) \d^2}{6}$.
The size of the impurities can be measured experimentally and is of the order
of
$20 \mu m$. We thus expect the correlation length to be a few times this size.
Its precise value depends on the details of the disorder.
For temperatures not
too close to transition temperature, $\th_{eq} \sim 20$ degrees, and
$\ga \sim 10^{19} Km^{-2}$.
This  leads to a typical  estimate $T_c \sim 10^{8} K$, in the experimental
conditions of
\cite{repain}.
Therefore $T/T_c$ is of order $10^{-8}$ and the system is effectively at
low temperatures,
justifying the low temperature limit in our computations.

\subsection{Comparison between experimental data and the theory neglecting
gravity}

We consider experimental data for height correlations
${\cal H}_{exp}(x)=\sqrt{\ol {<(\p(x)-\p(0))^2>}}$ for temperatures
$T=1.72, 1.8, 1.9, 1.93 K$.
The equilibrium angle $\th_{eq}$, the liquid-vapour interfacial tension $\ga$
and
the ``potential stength'' $W$ depend on the temperature, and so
the various experimental curves correspond {\it a-priori} to different values
of the
Larkin length. Instead, the correlation length $\d$ which depends only on the
substrate,
is expected to remain constant.
We have thus fitted the experimental curves to the theoretical
prediction in the absence of gravity (\ref{corr}) and (\ref{corr_grav}),
with the same $\d$ but different  $\xi$'s.

We proceed as follows. We first minimise the error function

\be\label{mini}
{\cal E}=\ds  \frac{1}{\ds \sum_j N_j}   \sum_{j=1}^{4} \sum_{i=1}^{N_j}
\biggl( \d {\cal H}(x_i/\xi_j)-{\cal H}_{exp}(x_i)\biggl)^2
\ee
\noi

\noi with respect to $\d$ and  $\xi_j$ for $j \in [1,4]$,
where $j$ denotes a given experimental curve at the temperature  $T_j$, $x_i$
the experimental points, $N_j$ the number of points of curve $j$, and
$\xi_j$ the correlation length at temperature $T_j$. This procedure yields
$\d \simeq 18 \mu m$ and ${\cal E} \simeq 0.3 \mu m ^2 $.

\begin{figure}[h,t]
$$
\epsfig{file=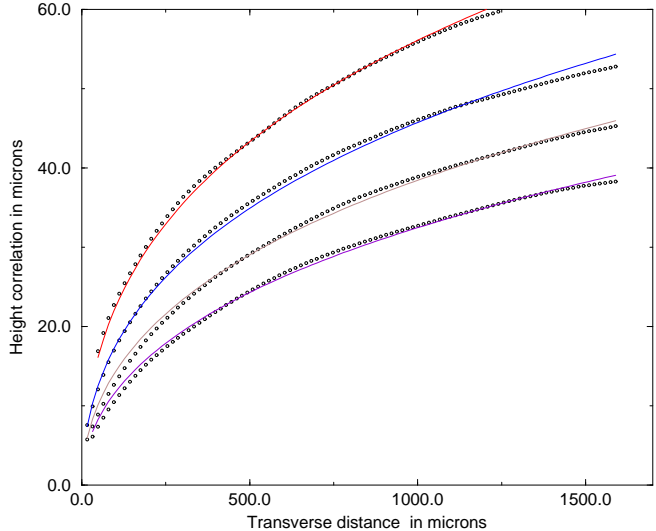,width=8.5cm,angle=-90}
$$
\caption{
The experimental data for the fluctuation of the line as a function of the
distance (circles).
{}From top to bottom, the temperatures are $1.93 K$,  $1.9 K$, $1.8 K$ and
$1.72 K$.
The full curves are the theoretical predictions in
absence of gravity (\ref{corr}), using the values of Larkin lengths shown
in table $1$.}
\label{fit2}
\end{figure}

In figure (\ref{fit2}), we show the four  experimental curves, together with
the corresponding theoretical fit $\Delta{\cal H} (x/\xi_j)$.
The values of the Larkin lengths $\xi_j$ are given in table $1$  below.
The dependance of $\xi$ on the temperature is related to the variations of
$W$, $\ga$ and $\th_{eq}$, which are not known well enough for a detailed
comparison between theory and experiment.
It should be noted that the determinations of $\Delta$ on the one hand, and the
$\xi_j$
on the other hand, are strongly correlated.
(To give an idea, with $\d=19 \mu m $, $\xi_1=60 \mu m $, $\xi_2= 104 \mu m$,
$\xi_3= 163 \mu m$, $\xi_4= 248 \mu m$
the error (\ref{mini}) differs from the previous case by about only $10 {\%}$).
In the absence of
detailed  information on the experimental error, it is thus difficult to give
an
error bar on $\xi $. On the other hand, the {\it ratios} of the Larkin lengths
in
different experiments are much less sensitive to this correlation.
They are given in table $1$ as well.

\begin{table}
\begin{center}
\begin{tabular}{|c|c|c|c|c|}
T(K)        &   1.72   &   1.8     & 1.9   & 1.93   \\
\hline
$\xi$($\mu m $)  &   220   &   145   & 92   & 53     \\
\hline
$\xi/\xi_1$     &   4.1  &    2.7    &     1.7    & 1   \\
\hline
\end{tabular}
\end{center}
\caption{Values of the Larkin length scale used in the fit to the experiment
shown
in figures (\ref{fit2},\ref{fit3}).}
\end{table}

\begin{figure}[h,t]
$$
\epsfig{file=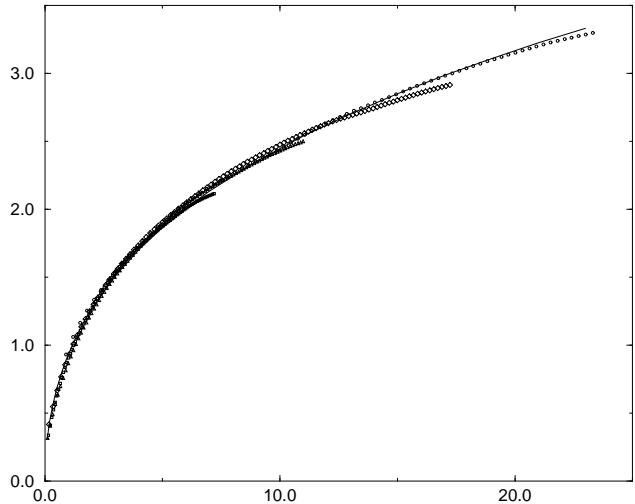,width=8.5cm,angle=-90}
$$
\caption{ Collapse plot of all the experimental data on fluctuations
of the line at various temperatures
onto the theoretical prediction neglecting gravity
(\ref{corr}), using the values of Larkin lengths shown in table $1$.}
\label{fit3}
\end{figure}

In figure (\ref{fit3}), we show a collapse plot of all the experimental curve
on  the
theoretical one  with gaussian correlated disorder.
We have rescaled  each experimental  curve by the corresponding
$1/\xi$ in the $x$ direction and by $1/\d$ in the $y$ direction. This collapse
gives
several interesting results: the curves nearly collapse one onto the other, as
expected
from the general form (\ref{corrgen}). Furthermore it seems that the simplest
correlation function
that we have studied in most details (\ref{function_f}) gives a reasonable
fit to the data. We have also checked that the small $x$ cut-off discussed in
section (\ref{sub_delta}) is irrelevant.
However there is also clearly, a systematic difference at large distances,
which we shall now discuss.

\subsection
{Comparison between experimental data and the theory including gravity}
In figure (\ref{fit3}), we note that each of the rescaled experimental curves
lies  below the theoretical curve at large distances.
Moreover   the experimental curves  have a slightly
larger curvature than the rescaled theoretical curve. These are
indications that gravity cannot be totally neglected. Indeed
the effective capillary length in the experimental conditions  is of the
order of  $2 mm$, and experimentally  the correlations are measured for
distances up to
about $1.5 mm$, which is actually not small compared with the capillary length.

To check whether gravity has or not a significant effect, we have tried to fit
the experiments with the full theoretical prediction including gravity
(\ref{corr_grav}).
Encouraged by the results of the
previous analysis, we keep to the case of a gaussian correlation function of
the disorder
given by (\ref{eq2}),(\ref{eq2bis}).
We have carried out the same analysis as in the previous subsection,
using as theoretical input ${\cal H}_g(x,\l)$ instead of ${\cal H}(x)$. The
capillary length
$L_c$ is not adjustable: it is calculated for the different temperatures from
the experimental
measurements of $\ga$, and are given in table $2$.
Therefore this new fit has the
same number of adjustable parameters as the previous one. We now minimise the
error function:

\be\label{mini_grav}
{\cal E}_g=\ds  \frac{1}{\ds \sum_j N_j} \sum_{j=1}^{4} \sum_{i=1}^{N_j}
\biggl( \d {\cal H}_g(x_i/\xi_j,\pi \xi_j/L_c^{j})-{\cal H}_{exp}(x_i)\biggl)^2
\ee
\noi As one could expect form the rather slow convergence of the
theoretical curves ${\cal H}_g(x,\l)$ towards the gravity-free one
${\cal H}(x)={\cal H}_g(x,0)$ at small $\lambda$ (see figure \ref{fit1}),
we find rather different values for the parameters. In this case,
$\d \simeq 75 \mu m$ and  ${\cal E}_g \simeq 0.26 \mu m^2$.
The  values of the  Larkin lengths for the different temperatures  are given in
table $2$.
We have the same problem of correlations between the determination of $\Delta$
and the
Larkin lengths as before. The ratios of the Larkin lengths in
different experiments are  also given in table $2$.

\begin{table}
\begin{center}
\begin{tabular}{|c|c|c|c|c|}
T(K)        &   1.72   &   1.8     & 1.9   & 1.93   \\
\hline
$L_c$ ($\mu m $)     &   1855  &  1838   &   1819   &   1823   \\
\hline
$\xi$($\mu m $)  &   480  &  425   &    365    &   295   \\
\hline
$\xi/\xi_1$     &  1.6  &    1.4    &     1.2    & 1   \\
\hline
\end{tabular}
\end{center}
\caption{Values of the experimental capillary lengths, and of the
Larkin length scale used in the fit to the experiment shown
in figure (\ref{fit4}).}
\end{table}

\begin{figure}[h,t]
$$
\epsfig{file=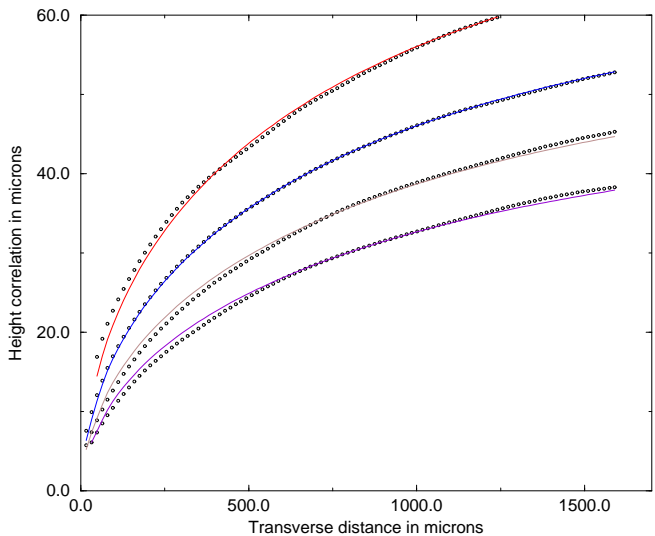,width=8.5cm,angle=-90}
$$
\caption{The experimental data for the fluctuation of the line
as a function of the distance (circles).
{}From top to bottom, the temperatures are $1.93 K$,  $1.9 K$, $1.8 K$ and
$1.72 K$.
The full curves are the  theoretical predictions in the
presence of gravity (\ref{corr_grav}), using the values of Larkin lengths shown
in table
$2$.}
\label{fit4}
\end{figure}

\noi In figure (\ref{fit4}), we show the experimental curves together
with  the
corresponding theoretical curve  ${\cal H}_g(x,\l)$ including gravity.
The fit is clearly better in this case since we have got rid of the systematic
drift from the theory for large $x$. The exponents of the two experimentally
observed regimes are  correctly predicted by a theory without gravity, but to
obtain
the correct values of the correlation length and Larkin lengths it is necessary
to
take into account the effect of gravity.

\section{Discussion and perspectives}
Our analytic computation for the wandering of a contact line on a disordered
substrate
fits quite well  the experimental data. We have shown that gravity effects are
far from
negligible and should be taken into account in order to extract the relevant
parameters
(which are basically the correlation length of the disorder as well as the
Larkin length scale)
from the experiments. Taking into account gravity in the theory clearly
improves the
fit to the data. Moreover, the theory including gravity will allow larger
experimental
length scales in the analysis.

A few comments  about the validity of our computation are the following. First
of all, we
have used in most of our analysis a simple form for the correlation function of
the
disorder (\ref{function_f}) which is not necessarily the correct one. It is
true that
some of our predictions are independant of this form, like for instance
the existence of a scaling behaviour in absence of gravity. It is an
experimental problem to have a better description of the pinning disorder at
work, and
we have shown that our computation can be extended to any type of correlation.
The simplest
one seems to give already a good account of the data. The variational method
which has been
used in this work is also an approximation. So far there is no other analytic
quantitative
method available, and it seems to work quite well, confirming previous evidence
found in other problems \cite{Engel, Kree,MPtoy}.

A more interesting question concerns our
assumption of an equilibrium situation.
We have  supposed that the line is in thermal equilibrium
in order to write the usual partition function, and at the end of the
calculation, we have taken the temperature to zero since, as we can see from
the numerical values of the parameters, thermal fluctuations are irrelevant.
 In doing so, we retain only the states with
the lowest energy. Experimentally the state of the line has no apparent reason
to be a low temperature  equilibrium state. It is difficult so far to
characterize
fully the metastable states
 that can
be reached dynamically by the experimental procedure. They might have
generically the same
statistical properties as the ground state, this is actually under study
(\cite{HMsuite}).
A less interesting but may be more realistic alternative, could be that because
of gravity, the
fluctuations of the line do not go much beyond the correlation length.  So,
even
though we are out of the Larkin regime, we are not yet deep in the random
manifold regime
and there are thus not so many metastable states.

A purely dynamical computation could also be done. While the properties of the
unpinned line
have already been studied in \cite{KE}, the dynamics of the pinned object could
also
be very interesting. If the correlation length of the disorder can be made
much smaller, so that thermal fluctuation are no longer negligeable, we expect
the onset of some  ageing
dynamics \cite{BCKM}. It will be interesting both to compute its properties
and to measure them.
In particular, this system might present  a nice situation to measure the
fluctuation-dissipation
ratio in a
system which has a full (continuous) replica symmetry breaking, offering a
chance to
measure directly the function $[\sigma](u)$.

\section*{Acknowledgements}
It is a pleasure to thank C. Guthmann, A. Prevost and  E. Rolley for several
enlightening exchanges and for giving us access to their data,
as well as  J.-P. Bouchaud and J.Vannimenus for useful discussions.

\newpage
\appendix
\section{Computation of the function $[\s] (u)$}
We give in this appendix some details of the calculation of the
function $[\s] (u)$. We have from section (\ref{S3}),

\be\label{eqA1}
\ds \s (u)=\ds \frac{2\b W}{\d^2}{\hat f'}\biggl(\frac{B(u)}{\d^2}\biggr) \\
\ee

\noi where

\be\label{eqA2}
B(u)=\frac{2}{\b}\int \frac{dk}{2\pi} ({\t g(k)}-g(k,u))
\ee

\noi and from \cite{MP},

\be\label{eqA3}
{\t g(k)}-g(k,u)=\frac{1}{u(c|k|+[\s ](u))}-\int_{u}^{1}\frac{dv}{v^2}
\frac{1}{c|k|+[\s ](v)}
\ee

\noi Differentiating (\ref{eqA1}) gives

\be\label{equation1}
\ds \s' (u)=\ds \frac{2\b W}{\d^4}B'(u)
{\hat f''}\biggl(\frac{B(u)}{\d^2}\biggr) \\
\ee

\noi and replacing $B'(u)$ in (\ref{equation1}) by its expression

\be\label{eqA4}
\ds B'(u)=\ds - \frac{2}{\b \pi c}\frac{\s '(u)}{[\s ](u)}
\ee

\noi leads to

\be\label{eqA5}
\begin{array}{c}
\ds \s' (u) =  0 \quad  \quad  \mbox{or} \\
\\
\ds 1 = \ds -\frac{4W}{\pi c \d^4}\frac{1}{[\s] (u)}
{\hat f}''\biggl(\frac{B(u)}{\d^2}\biggr)
\end{array}
\ee

\noi We express $\ds \frac{B(u)}{\d^2}$ in terms of $[\s ](u)$  by inverting
the
second equation of (\ref{eqA5}). This gives:

\be\label{eqA6}
\ds {\hat f}''^{-1}\biggl(\ds -\frac{\pi c \d^4}{4W}
[\s](u)\biggr)=\ds \frac{B(u)}{\d^2}
\ee

\noi Differentiating (\ref{eqA6}), and using expression (\ref{equation1})
to express $B'(u)$, and the fact  that $[\s ]'(u)=u\s '(u)$, we get

\be\label{eqA7}
\ds \frac{{\hat f}^{'''}}{{\hat f}^{''}}
\biggl(\frac{B(u)}{\d^2}\biggr)=\ds
-\frac{\b \pi c \d^2}{2}u=\ds -\frac{3T_c}{2T}u
\ee

\noi where

\be\label{eqA8}
{\hat f}'(x)=\frac{1}{2\sqrt{1+x}}
\ee

\noi From (\ref{eqA7}) and (\ref{eqA8}), we have

\be\label{eqA9}
\frac{B(u)}{\d^2}=-1+\frac{3T}{\pi c u \d^2}=-1+\frac{T}{T_c}\frac{1}{u}
\ee

\noi Multiplying both sides of equation
(\ref{eqA4}) by $u$,  and using (\ref{eqA9}) we get after integrating over u

\be\label{eqA10}
[\s](u)=A u^{3/2}
\ee

\noi The breakpoint $u_c$, above which the solutions to (\ref{eqA9}) and
(\ref{eqA10})
are no longer valid is given by

\be\label{eqA11}
B(u_c)=\d^2 \biggl(-1+\frac{T}{T_c}\frac{1}{u_c}\biggr)
   =\frac{2}{3}\frac{T}{T_c}\ln \biggl(1+\frac{2\pi c}{\d [\s](u_c)}\biggr)
\ee

\noi When $\ds \frac{T}{T_c}$ goes to $0$,
$u_c \simeq\ds \frac{T}{T_c}$. Now since $B(u)$ tends to infinity
when $u$ tends to $0$,  we have $\s (0)=0$.
For $u \geq  u_c$, $B(u)=B(u_c)$ and $[\s](u)=[\s](u_c)$.
To obtain $A$, we can differentiate (\ref{eqA10}) and compare the result
with the expression
for $\s '(u)$. We find $\ds A=\frac{W}{\pi c \d^4}\frac{1}{u_c^{3/2}}$.
For $u \geq u_c$,
 $\s '(u)=0$ and so $\ds [\s](u)=[\s](u_c)=\ds \frac{W}{\pi c \d^4}$.

\section{General form of the correlation function for arbitrary disorder}

In this appendix we derive the height correlation function for
a more general form of the function $f$ appearing in the correlation
function of the  disorder (\ref{eq2}). We only impose that $f(|u|) \sim \l
\sqrt{|u|}$
for large $u$ where $\l$ is some constant,
such that ${\hat f}(u) \sim  \sqrt{|u|}$.
We shall keep the same notations as in appendix {\bf A}.
In this more general case, equations (\ref{eqA1}),(\ref{eqA4}),(\ref{eqA7})
from
appendix {\bf A} are still valid. We define the function $h^{-1}$ for positive
$x$
as

\be\label{eqA12}
h^{-1}(x)=\ds \frac{{\hat f}^{'''}(x)}{{\hat f}^{''}(x)}
\ee

\noi where $\ds h^{-1}(x) \sim -\frac{3}{2x}$ for large $x$. The asymptotic
behaviour
of $h(y)$ for small and negative $y$ is then $\ds -\frac{3}{2y}$.
We now express $B(u)$ in terms of $h$, from equations (\ref{eqA7}) and
(\ref{eqA12}).
This gives

\be\label{eqA13}
B(u)=\d ^2 h \biggl(- \frac{\pi c \d ^2}{2T}u \biggr)
\ee

\noi Differentiating the previous equation (\ref{eqA13}) gives

\be\label{eqA14}
B'(u)=-\frac{\pi c \d ^4}{2T} h' \biggl(- \frac{\pi c \d ^2}{2T}u \biggr)
\ee

\noi which can be rewritten as

\be\label{eqA15}
\frac{2T}{3T_c} \frac{d}{du} \log[\s](u)=w h'(-w)
\ee

\noi where $w=\ds \frac{\pi c \d ^2}{2T}u$. Integrating (\ref{eqA15}) gives

\be\label{eqA16}
\ds [\s ](u)=\ds [\s ](\ep)
\ds \exp \ds \int_{3\ep/2u_c}^{3u/2u_c}dw \  w h'(-w)
\ee

\noi  For $\ep \leq  u \ll u_c$, we can use the asymptotic form of
$h$ in (\ref{eqA16}), which  then reads

\be\label{eqA17}
[\s ](u)=[\s ](\ep)\biggl(\frac{u}{\ep}\biggr)^{3/2}
\ee

\noi Now for small $u$, equation (\ref{equation1}) becomes in this case

\be\label{eqA18}
\s'(u)= \frac{3}{2}\frac{W}{\pi c \d ^4}
\biggl(\frac{T_c}{T}\biggr)^{3/2}\frac{1}{\sqrt{u}}
\ee

\noi and so for small $u$, since $[\s](0)=0$, we get

\be\label{eqA19}
[\s](u)= \frac{W}{\pi c \d ^4}\biggl(\frac{u}{u_c}\biggr)^{3/2}
\ee

\noi A comparison of  this last expression with (\ref{eqA17}) gives

\be\label{eqA20}
[\s](\ep)= \frac{W}{\pi c \d ^4}\biggl(\frac{\ep}{u_c}\biggr)^{3/2}
\ee

\noi Replacing  this last expression in (\ref{eqA16}) and taking $\ep$ to zero
leads to

\be\label{eqA21}
[\s](u)=\frac{W}{\pi c \d ^4} {\cal S}(u)
\ee

\noi where

\be\label{eqA22}
{\cal S}(u)=\ds \biggl(\frac{u}{u_c}\biggr)^{3/2}
\exp \ds {\int_{0}^{3u/2u_c} dw
\ w \biggl(h'(-w)-\frac{3}{2w^2}\biggr)}
\ee

\noi For the sake of simplicity, we will suppose that in this case the
break-point up to which expression  (\ref{eqA22}) is valid, is also $u_c$
in the limit of low temperatures. This implicitely requires that $h(-3/2)=0$.
Then for $u \geq u_c$

\be\label{eqA22bis}
{\cal S}(u)={\cal S}(u_c)=\ds
\exp \ds {\int_{0}^{3/2} dw
\ w \biggl(h'(-w)-\frac{3}{2w^2}\biggr)}
\ee

\noi When ${\hat f'}$ has the simple form (\ref{eqA8}),
$\ds h'(-w)=\frac{3}{2w^2}$, and  we recover the expression
of $[\s](u)$ derived in appendix {\bf A}.In the limit of low temperatures
and for $\d \ll |x-x'| \ll L$, the  height correlation function is given by

\be
\sqrt{\ol {<(\p(x)-\p(x'))^2>}} =\d {\cal G}\biggl(\frac{x-x'}{\xi}\biggr)
\ee

\noi with

\be
\begin{array}{ll}
{\cal G}^2(x)=& \ds \frac{4}{3} \biggl(\frac{x}{\pi}\biggr)^{2/3}
\int_{0}^{\infty}dk
\frac{(1-\cos(k))}{k^{5/3}} \int_{0}^{(x/{\pi k})^{1/3}}dv \
\ds \frac{1}{v^3+\exp \ds  \biggl(
-\int_{0}^{\frac{3v^2}{2}(\pi k/x)^{2/3}} dw \ds
\biggl(h'(-w)-\frac{3}{2w^2}\biggr)\biggr)}

\\ \\

& \quad \quad \quad \quad +\ \ds   \frac{2x}{3\pi} \int_{0}^{\infty}dk
\ds \frac{(1-\cos(k))}{k\biggl(x/{\pi}+k\exp \ds  \biggl(
-\int_{0}^{3/2} dw \ds  \biggl(h'(-w)-\frac{3}{2w^2}\biggr)\biggr)\biggr)}
\end{array}
\ee

\section{Effect of  gravity}

In this appendix, we consider a specific case of the disorder given by
equations (\ref{eq2}) and (\ref{eq2bis}).
To take into account gravity, we must replace the kernel $|k|$ by
$\sqrt{|k|^2+\mu ^2}$,
with $\ds \mu =\frac{1}{L_c}$, where $L_c$ is the capillary length.
The equations derived in appendix {\bf A} are thus no longer valid.
If $\s'(u)$ is not zero then, equation (\ref{eqA6}) of appendix {\bf A} becomes

\be\label{eqC1}
\frac{B(u)}{\d ^2}={\hat f}''^{-1}
\biggl(-\frac{\pi c \d ^4}{4W}{\cal K}([\s](u))\biggl)
\ee

\noi where

\be\label{eqC2bis}
{\cal K}(x)=c \mu \ g \biggl(\frac{x}{c \mu}\biggr)
\ee

\noi with

\be\label{eqC2}
\frac{1}{g(x)}= \int_{0}^{\infty}
\frac{dk}{(\sqrt{k^2+1}+x)^2}
\ee

\noi Differentiating (\ref{eqA3}) gives

\be\label{eqC3}
B'(u)=-\frac{2\s '(u)}{\pi \b c} \frac{1}{{\cal K}([\s](u))}
\ee

\noi Differentiating (\ref{eqC1}), and using (\ref{equation1}) and
(\ref{eqA8}),
we have

\be\label{eqC4}
1+\frac{B(u)}{\d ^2}= \frac{T}{T_c} \frac{1}{u{\cal K}([\s](u))}
\ee

\noi Differentiating the previous expression (\ref{eqC4}), and using
(\ref{eqC3}),
we can express $[\s](u)$ as

\be\label{eqC5}
\frac {{\cal K}([\s](u))}{({\cal K}'([\s](u)))^{3/2}}=Au^{3/2}
\ee

\noi where $A$ is a constant to be determined. Replacing in equation
(\ref{equation1}),
$B'(u)$ by its expression (\ref{eqC1}) and reexpressing (\ref{eqA8}) using
(\ref{eqC4})
leads to

\be\label{eqC6}
1=\frac{W}{\pi c \d ^4} \frac{1}{u_c^{3/2}}
\frac{(u{\cal K}'([\s](u)))^{3/2}}{{\cal K}([\s](u))}
\ee

\noi Comparing the previous expression with (\ref{eqC5}) gives

\be\label{eqC7}
A=\frac{W}{\pi c \d ^4} \frac{1}{u_c^{3/2}}
\ee

\noi To express $[\s]$, it is convenient to introduce the function ${\cal I}$

\be\label{eqC8}
{\cal I}(x)=\ds \frac{g(x)}{(g'(x))^{3/2}}
\ee

\noi which is strictly positive and increasing.  ${\cal I}(0) \simeq 0.87$
and ${\cal I}(x)$ goes as $x$ when $x$ goes to infinity.

\noi The inverse function ${\cal I}^{-1}$ is thus defined on the interval
$[{\cal I}(0),\infty]$. Since $[\s](u)$ must be continuous and  $[\s](0)=0$,
the function $[\s]$ necessarily has a first plateau where $[\s](u)=0$
from $u=0$ up to a  value $u_1$ given by

\be\label{eqC9}
{\cal I}(0)=\frac{W}{\pi c \d ^4} \frac{1}{\mu c}
\biggl(\frac{u_1}{u_c}\biggr)^{3/2}
=\frac{L_c}{\pi \xi} \biggl(\frac{u_1}{u_c}\biggr)^{3/2}
\ee

\noi For $u_1 \leq u \leq u'_c$, where $u'_c$ is the new break-point to be
determined

\be\label{eqC10}
[\s](u)=c \mu \ {\cal I}^{-1}
\biggl(\frac {L_c}{\pi \xi}\biggl(\frac{u_1}{u_c}\biggr)^{3/2}\biggr)
\ee

\noi The break-point $u'_c$ is obtained using (\ref{eqC4}) and (\ref{eqA3})
is given

\be\label{eqC11}
\frac{B(u'_c)}{\d ^2}=-1+u_c \frac{1}{u'_c{\cal K}'([\s](u'_c))}
=\ds \frac{2}{3}u_c \int \ds \frac{dk}
{\ds \biggl(\sqrt{k^2+1}+\frac{[\s](u'_c)}{\mu c}\biggr)}
\ee

\noi and when $T$ goes to zero

\be\label{eqC10bis}
u_c \simeq u'_c{\cal K}'([\s](u'_c)) \
= \ds u'_c g' \biggl(\frac{[\s](u'_c)}{\mu c}\biggr)
\ee

\noi where

\be\label{eqC12}
\ds g'(u)=\ds 2 \int_0^{\infty}  \frac{dq}{(\sqrt{q^2+1}+u)^3}\
\ds \biggl(\int_0^{\infty} \frac{dq}{(\sqrt{q^2+1}+u)^2}\biggr)^{-2}
\ee

\noi Since $g'$ is a strictly increasing function, with  $g'(0)=8/\pi ^2$ and
$g'(\infty)=1$, in the limit of low temperatures
$\ds u_c \leq u'_c \leq \frac{u_c}{g'(0)}$.
Since ${\cal I}(x)$ is almost linear, we can suppose that
$\ds [\s](u'_c) \simeq \frac{W}{\pi c \d ^4}
\biggl(\frac{u'_c}{u_c}\biggr)^{3/2}$
and so equation (\ref{eqC10bis}) can be rewritten as

\be\label{eqC13}
\frac{1}{\l ^{2/3}}=\Omega ^{2/3} g'(\Omega)
\ee

\noi with

\be\label{eqC14}
\Omega=\frac{1}{\l} \biggl(\frac{u'_c}{u_c}\biggr)^{3/2} \quad \mbox{and} \quad
\l=\frac{\pi  \xi}{L_c}
\ee

\noi We can solve for $u'_c$ perturbatively using the solution without gravity.
As a first approximation, we can take for $\l$ the its value in the absence
of gravity. We can then solve numerically for $\Omega$ and for $u'_c$.
For our experimental data, we get by this method $u'_c \simeq u_c$.
For $u \geq u'_c$, $[\s](u)=[\s](u'_c)$.
The  height correlation  function is then given by

\be\label{eqC15}
\sqrt{\ol {<(\p(x)-\p(x'))^2>}} =\d{\cal H}_g\biggl(\frac{x-x'}{\xi},
\frac{\pi  \xi}{L_c} \biggr)
\ee

\noi where

\be
\begin{array}{ll}
{\cal H}_g^2 (x,\l)=&\ds \frac{4}{3} \int_{0}^{\infty} dk
\frac{(1-\cos(kx/\pi))}{\sqrt{k^2+\l^2}}
\ds \int_{(\l {\cal I}(0))^{1/3}}^{(\l \Omega)^{1/3}}\frac{dw}{w^3}
\frac{\l {\cal I}^{-1}(w^3/\l)}{\l {\cal I}^{-1}(w^3/\l)+\sqrt{k^2+\l^2}}

\\ \\

& + \ds \frac{2}{3}  (\l \Omega)^{-2/3}
\int_{0}^{\infty} dk \ \frac{(1-\cos(kx/\pi))}{\sqrt{k^2+\l^2}}
\frac{\l {\cal I}^{-1}(1/\l)}{\l {\cal I}^{-1}(1/\l)+\sqrt{k^2+\l^2}}
\end{array}
\ee
\end{document}